\documentclass[apjl,twocolumn]{openjournal}
\usepackage{amsmath}
\usepackage{booktabs}
\usepackage{multirow}
\usepackage{color}
\usepackage{soul}
\usepackage{upgreek}

\usepackage[dvipsnames]{xcolor} 

\usepackage[breaklinks,colorlinks,citecolor=blue,urlcolor=blue]{hyperref}
\usepackage{orcidlink}

\renewcommand{\arraystretch}{1.2}  
\usepackage{listings}
\usepackage{color}
\definecolor{dkgreen}{rgb}{0,0.6,0}
\definecolor{gray}{rgb}{0.5,0.5,0.5}
\definecolor{mauve}{rgb}{0.58,0,0.82}
\definecolor{golden}{rgb}{0.86,0.65,0.01}
\begin{document}

\title{Does God play dice with star clusters? \vspace{-1.5cm}}

\author{Michael Y. Grudi\'{c}$^{1,\dagger}$\orcidlink{0000-0002-1655-5604}}
\author{Stella S. R. Offner$^{2}$\orcidlink{0000-0003-1252-9916}}
\author{D\'avid Guszejnov$^{3,\dagger}$\orcidlink{0000-0001-5541-3150}}
\author{Claude-Andr{\'e} Faucher-Gigu{\`e}re$^{4}$\orcidlink{0000-0002-4900-6628}}
\author{Philip F. Hopkins$^{5}$\orcidlink{0000-0003-3729-1684}}

\affiliation{$^1$Carnegie Observatories, 813 Santa Barbara St, Pasadena, CA 91101, USA}
\affiliation{$^{2}$Department of Astronomy, The University of Texas at Austin, TX 78712, USA}
\affiliation{$^3$Center for Astrophysics $|$ Harvard \& Smithsonian, 60 Garden Street, Cambridge, MA 02138, USA}
\affiliation{$^{4}$CIERA and Department of Physics and Astronomy, Northwestern University, 1800 Sherman Ave, Evanston, IL 60201, USA}
\affiliation{$^{5}$TAPIR, Mailcode 350-17, California Institute of Technology, Pasadena, CA 91125, USA}

\thanks{$\dagger$NASA Hubble Fellow}
\email{Corresponding author: mgrudic@carnegiescience.edu}

\begin{abstract}
When a detailed model of a stellar population is unavailable, it is most common to assume that stellar masses are independently and identically distributed according to some distribution: the universal initial mass function (IMF). However, stellar masses resulting from causal, long-ranged physics cannot be truly random and independent, and the IMF may vary with environment. To compare stochastic sampling with a physical model, we run a suite of 100 {\small STARFORGE} radiation magnetohydrodynamics simulations of low-mass star cluster formation in $2000M_\odot$ clouds that form $\sim 200$ stars each on average. The stacked IMF from the simulated clouds has a sharp truncation at $\sim 28 M_\odot$, well below the typically-assumed maximum stellar mass $M_{\rm up} \sim 100-150M_\odot$ and the total cluster mass. The sequence of star formation is not totally random: massive stars tend to start accreting sooner and finish later than the average star. However, final cluster properties such as maximum stellar mass and total luminosity have a similar amount of cloud-to-cloud scatter to random sampling. Therefore stochastic sampling does not generally model the stellar demographics of a star cluster as it is forming, but may describe the end result fairly well, if the correct IMF -- and its environment-dependent upper cutoff -- are known.

\end{abstract}

\maketitle

\section{Introduction}

Astrophysicists aim to interpret light from galaxies and star clusters in terms of their stellar populations and the physical processes that affect them. More often than not, detailed knowledge of the individual stars is not available, so inferences about the system rely on a model of their stellar demographics. The cornerstone of such models is the stellar initial mass function (IMF): the distribution of stellar birth masses \citep{salpeter_slope, Hopkins_A_2018_IMF_obs_review}. The standard approach to modeling a galaxy or star cluster is to sample stellar masses from the IMF and evolve the stars forward in time according to a stellar evolution model \citep[e.g.][]{starburst99,bruzual:2003.ssp}. 

A common model assumption is that stellar masses are sampled independently and identically from an invariant IMF (``stochastic sampling"), i.e., the probability of a star forming with mass $m_{\rm 1}$ does not depend on the mass of another star $m_{\rm 2}$, nor the macroscopic properties of the system the star is forming in. Or, in other words, 10 little star clusters are statistically identical to one big one. This is the simplest approach and is motivated by the observed lack of strong, systematic variations in the IMF in normal star-forming environments \citep{bastian:2010.imf.universality,offner:2014.imf.review}. 

Stochastic sampling 
should not necessarily be interpreted physically: the initial gas flows in molecular clouds that produce individual stars cannot be truly random and independent of each other. The formation of one star can affect another via gravity and feedback in the form of protostellar outflows, winds, radiation, supernovae, and cosmic rays. So stochastic sampling cannot be the ground-truth star formation process, and if star clusters deviate from it significantly then so will any resulting inferences from the integrated light of galaxies \citep{weidner.kroupa:2006,kroupa_2013_imf_review,kroupa:2021.imf}.

Of course, {\it all} models are wrong at some level, but some are useful. 
Stochastic sampling from an invariant IMF has enjoyed wide application due to its simplicity and the lack of an alternative model with compelling physical motivation. It has been used to characterize star clusters and their demographics from photometry \citep{fumagalli:2011.imf,slug}, where it gives a marginally-better fit to the data than IMF-averaged photometric models without stochasticity \citep{krumholz:2015.slug} and can enable inferences about the highly-incomplete low-mass cluster population, if accurate \citep{2023arXiv230105912T}. Stochastic sampling has also seen increasing use in high-resolution galaxy and molecular cloud simulations \citep{fujii:2015.cluster.formation,hu:2017.imf.sampling, su:2017.discreteness,lahen:2020.griffin.sims, smith:2020.galaxy.imf.sampling,liow.rieder:2022.imf.sampling}. On molecular cloud ($\lesssim 100 \rm pc$) scales, the granularity of feedback and gravity from individual stars is quite important; the specific timing, sampling, and distribution of massive star formation has a large effect on subsequent star formation \citep{elephant,dinnbier.walch:2020.imf.sampling.feedback,lewis:2022.torch.massivestars}. However, 
lacking a detailed physical picture of star cluster formation, it has been difficult to assess the fidelity of stochastic sampling fully.

Recently, the {\small STARFORGE} framework\footnote{\url{https://www.starforge.space}} \citep{starforge.methods} has enabled radiation magnetohydrodynamics (RMHD) star formation simulations with significantly enhanced realism and self-consistency. By accounting for key feedback processes and resolving individual star formation, the simulations can follow the entire sequence of star formation from initial cloud collapse, through star formation and accretion, feedback, to the eventual cloud disruption \citep{starforge.fullphysics}. This allows us to predict the final masses of stars and hence the IMF in a consistent framework \citep{guszejnov:2022.starforge.imf}.

In this paper we present the results of a statistical ensemble of 100 {\small STARFORGE} simulations and measure their combined and individual stellar mass statistics. The statistical power of this suite allows us to compare stochastic sampling models to a physical model of star cluster formation with all relevant processes for the first time. In particular we aim to test whether statistical sampling, independent of cloud mass, provides a good representation of star formation process realized in nature. We will focus on the statistics of massive ($\gtrsim 8 M_\odot$) stars, which dominate the light and stellar feedback from young star clusters, and hence are most relevant for photometric modeling and numerical simulations.

\section{Simulations}
We use the {\small GIZMO} code \citep{hopkins:gizmo} to run an ensemble of 100 {\small STARFORGE} RMHD star cluster formation simulations. 
The simulations use the ``full" {\small STARFORGE} physics setup introduced in \citet{starforge.fullphysics}, with numerical methods for sink particles, stellar evolution, and feedback coupling described in full in \citet{starforge.methods}. The simulations solve the equations of ideal magnetohydrodynamics (MHD) using the {\small GIZMO} Meshless Finite Mass MHD solver \citep{hopkins:gizmo.mhd} with constrained-gradient divergence minimization \citep{hopkins:gizmo.mhd.cg} and a fixed Lagrangian mass resolution of $10^{-3} M_\odot$.

Gas cells turn into sink particles if they satisfy all of a list of criteria intended to identify centers of runaway gas collapse. They then accrete other gas cells that have sufficient density, proximity to the sink, low angular momentum, and gravitational boundedness \citep[e.g.][]{Bate_1995_accretion}. There is not stochasticity in these criteria. Each sink particle hosts a star that accretes the sink's gas reservoir smoothly, evolving in radius and deuterium abundance according to the \citet{Offner_2009_radiative_sim} protostellar evolution model. Stars on the main sequence may still accrete and move along it, and their bolometric luminosity and radius are given by the \citet{tout_1996_mass_lum} fits. Stars' emergent spectra are approximated as a black-body with $T_{\rm eff}=5780\rm K (L_\star/L_\odot)^{1/4}(R_\star/R_\odot)^{-1/2}$, and their metallicity- and luminosity-dependent mass-loss rate and wind velocity are given by the fitting formulae in \citet{starforge.methods,starforge.fullphysics}. Massive stars can progress to a Wolf-Rayet phase with strong winds and an eventual supernova, but this does not occur during the star-forming phase of any of the clouds simulated here, which are disrupted primarily by protostellar outflows and ionizing radiation. 

We solve the time-dependent, frequency-integrated radiative transfer (RT) equation with the {\small GIZMO} meshless frequency-integrated M1 solver \citep{hopkins.grudic:2018.rp,hopkins:2020.fire.rt} with a reduced speed of light of $30\,\rm km\,s^{-1}$, in 5 frequency bands ranging from Lyman continuum to far IR. The different bands couple to matter through photoionization, photodissociation, photoelectric heating, and dust absorption, accounting for photon momentum. We account for radiative cooling and heating from all major molecular, atomic, nebular, and continuum processes using the cooling module shared with the {\small FIRE-3} simulations \citep{fire3}, except that we perform dust radiative transfer self-consistently with the IR band of the RT solver, evolving the dust, gas, and radiation temperatures independently.

\subsection{Initial conditions} 
All 100 simulations start with a uniform-density molecular cloud of mass $M=2000M_\odot$ and radius $R=3\rm pc$, in a $30\rm pc$ periodic box filled with an ambient medium $10^{3}$ times less dense. The initial magnetic field is uniform and normalized to give a mass-to-flux ratio of 4.2 times the critical value inside the cloud. The simulations differ only in their initial random velocity fields, which are different statistical realizations of a Gaussian random field with a $\propto k^{-2}$ power spectrum, all normalized so that the initial virial parameter is $\alpha_{\rm turb} = 5 \sigma^2 R/(3 G M) = 2$, with a natural mix of compressive and solenoidal modes. We use a fixed quasi-Lagrangian mass resolution of $10^{-3}M_\odot$ for normal gas cells and $10^{-4}M_\odot$ for protostellar jet and stellar wind cells, hence each cloud initially consists of $2\times 10^6$ gas cells.

\section{Results}
All simulations are run for $10\, \rm Myr$ (5 cloud freefall times) and follow the usual qualitative sequence predicted by molecular cloud simulations with star formation and feedback, e.g. as documented in detail in \citet{starforge.fullphysics}. Turbulent motions seed dense substructures that go on to collapse into stars, which then proceed to accrete and disperse the cloud via feedback, eventually halting star formation. The clouds form a total of 19,009 stars totaling $1.26\times 10^4 M_\odot$, for an average final cluster mass of $126 M_\odot$, i.e. a final star formation efficiency of 6\%, broadly consistent with cloud-scale SFEs inferred in the Solar neighborhood \citep{evans:2009.sfe,evans:2014.sfe,2016AJ....151....5M}.  Roughly 80\% of the stellar systems appear to be expanding freely at the end of their respective simulations, and the others have a steady half-mass radius and are likely bound and destined to dissolve over a longer timescale. For the purposes of this paper we adopt a broad definition of ``cluster" and do not distinguish between expanding associations and bound clusters. 

\begin{figure}
    \centering
    \includegraphics[width=\columnwidth]{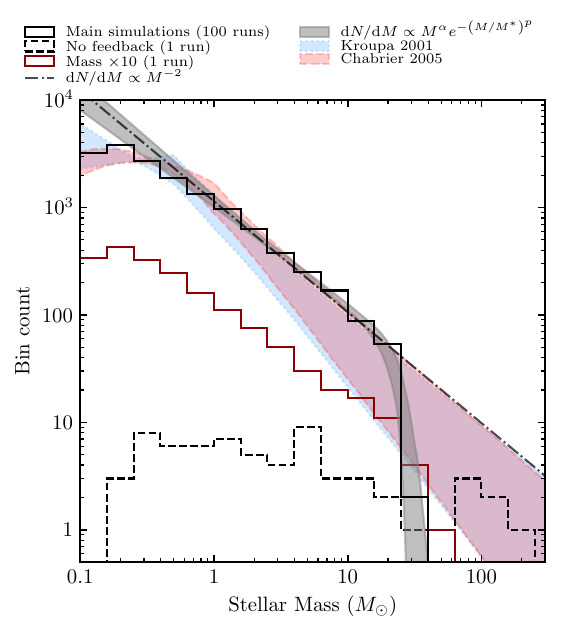}
    \caption{Histogram of the 19,009 ZAMS stellar masses resulting from the stacked 100 star cluster formation simulations. For comparison we plot estimated $\pm \sigma$ probable ranges of the \citet{kroupa:imf} and \citet{chabrier_imf} Milky Way field IMFs, using parameter uncertainties quoted in \citet{kroupa:imf} and assuming $\log m_{\rm c}/M_\odot=\log 0.2\pm 0.1 \rm dex$ and $\sigma=0.55\pm0.10 \rm dex$ for the \citet{chabrier_imf} form, controlling all IMFs for the number of $>0.1M_\odot$ stars. The dashed line plots the best-fit $\propto M^{-2}$ power-law, and the shaded region plots the $\pm 3 \sigma$ confidence region of a modified Schechter function fit, sampled from the posterior distribution of the model parameters given the $>1M_\odot$ sample. The two lower histograms show the IMF obtained from one $2\times 10^4 M_\odot$ cloud run with the same physics model, and a run with the same initial conditions but no feedback.}
    \label{fig:total_imf}
\end{figure}

The spectrum of zero-age main sequence masses of stars formed in all 100 simulations is plotted in Figure \ref{fig:total_imf}, with the \citet{kroupa:imf} and \citet{chabrier_imf} Solar neighborhood field star IMFs for comparison. This IMF is similar to the one that we have presented for clouds with these parameters and physics in previous work \citep{starforge.fullphysics,guszejnov:2022.starforge.imf}, with a power-law slope of $\sim -2$ in the $1-10M_\odot$ range and a turnover or break around $0.2 M_\odot$. The median stellar mass is $0.22 M_\odot$ and the log-dispersion is $\sigma_{\mathrm{log} M} = 0.52\,\rm dex$, similar to estimates for the Solar neighborhood field population \citep{chabrier_imf}. The high-mass slope $\alpha=-2$ is shallower than the normally-assumed -2.35, but well within the variation measured between individual clusters and within the uncertain range quoted for the Milky Way field population \citep{kroupa:imf}.

\subsection{The high-mass IMF truncation}
\label{sec:imf_truncation}
Stacking 100 realizations unveils a new feature of the IMF: a high-mass truncation. 
Figure \ref{fig:total_imf} shows that there is an unambiguous high-mass break around the sample maximum stellar mass of $28 \rm M_\odot$, for the 100 low-mass clusters\footnote{For our average cluster mass of $M_{\rm ecl}=126M_\odot$, our maximum stellar mass $M_{\rm max}=28M_\odot$ lies a factor of $\sim 2$ above the observed data points in the $M_{\rm ecl}-M_{\rm max}$ compilation presented in \citet{2023A&A...670A.151Y}.}. To assess this statistically, note that if the $\propto M^{-2}$ power-law portion extrapolated to $M_{\rm up} = 150 M_\odot$, we would expect $\sim$75 stars more massive than $28 M_\odot$ in our sample. The probability of obtaining zero stars above the cutoff is $2\times 10^{-33}$ assuming Poisson statistics. Even a steeper $\propto M^{-2.35}$ power-law up to $150M_\odot$ can be discarded with $p=5\times10^{-12}$ by this argument. The maximum stellar mass found in these simulations is confidently inconsistent with the size-of-sample effect from sampling a pure power-law extending to the largest-known stellar masses \citep[e.g.][]{Crowther_most_massive_stars_R136_2016}, a standard model assumption. The sample maximum is also significantly smaller than even the smallest cluster mass in the sample ($\sim 80M_\odot$), so it is not directly explained by the limited mass budget.


There is no physical reason to expect a perfectly-sharp truncation, so we model the high-mass ($>1M_\odot$) IMF as a power-law with a smooth truncation factor. A modified Schechter function with varying sharpness 
\begin{equation}
    \mathrm{d}N/\mathrm{d}M \propto M^{\alpha} \exp\left(-\left(M/M^\ast\right)^p\right)
\end{equation}
provides a better fit than the standard \citet{schechter:1976} form with $p=1$, e.g., its additional parameter is strongly preferred by the \citet{schwarz:1978.bic} information criterion ($\Delta \mathrm{BIC} = -17.5$). Assuming flat priors with $\alpha \in [0,-4]$, $\log M^\ast \in [0,4]$, and $p\in [0,10]$ the posterior distribution gives marginalized parameters $\alpha = -2.00\pm 0.02$, $M^\ast=24\pm 4M_\odot$, and $p=7 \pm 2$, and we plot the $3\sigma$ confidence interval of the parameter distribution in Figure \ref{fig:total_imf}.

This sample IMF is also statistically distinct from those found in various previous simulations using the same setup with different initial conditions \citep{starforge.fullphysics,guszejnov:2022.starforge.imf}. Most of those exhibited the same high-mass slope of -2 but produced more-massive stars in clusters with fewer members. For example, in Figure \ref{fig:total_imf} we also plot the IMF from a cloud run with an identical physics setup, surface density, virial parameter, and mass-to-flux ratio, but 10 times greater mass ($2\times 10^4 M_\odot$). This single cloud produces about $10\times$ fewer stars than the 100 smaller clouds. Despite this, the massive cloud produces 3 stars more massive than the $28 M_\odot$ maximum of the smaller clouds, with a maximum of $44M_\odot$. This strongly suggests that the upper cutoff predicted by the {\small STARFORGE} models scales in some way on the properties of the host GMC, in agreement with other recent simulations \citep{2023MNRAS.525.6182S}. We do emphasize that our model space can still account for the existence of $\gtrsim 100 M_\odot$ stars - e.g. the $10 \times$ higher surface-density cloud in \citet{guszejnov:2022.starforge.imf} formed a $107M_\odot$ star.   

Figure \ref{fig:total_imf} also shows the IMF from a $2000M_\odot$ cloud with the same bulk properties as the others, but evolved without any stellar feedback physics. Despite forming only 69 stars, this cloud produced 8 stars more massive than the most massive star in the feedback simulations, with a maximum stellar mass of $224M_\odot$. Clearly then the upper cutoff is highly sensitive to feedback. And because feedback generally operates with variable effectiveness depending on the environment \citep{fall:2010.sf.eff.vs.surfacedensity}, the upper cutoff is naturally sensitive to cloud properties.
 
Due to this variation in the upper truncation in the IMF, no sampling model employing the usual constant $\sim 100-150 M_\odot$ upper cutoff -- stochastic or not -- adequately describes the statistics of the simulated star clusters. Instead, a model in which the truncation depends on the environment is required.

\subsection{Sequence of massive star formation versus random sampling}
\label{sec:sfsequence}
We will now compare the number of massive ($>8M_\odot$) stars in the clusters at a given time with the mean number expected, $\langle N_{\rm massive} \rangle = f_{\rm massive} N_{\rm cl}$, where $f_{\rm massive}=0.012$ the overall fraction of $>8M_\odot$ stars in the sample IMF. We count massive stars in two different ways: (1) we include a protostar in the count after its formation if it will {\it eventually} be massive, and (2) we include a star in the count only after it has reached $8M_\odot$.

\begin{figure}
    \centering
    \includegraphics[width=\columnwidth]{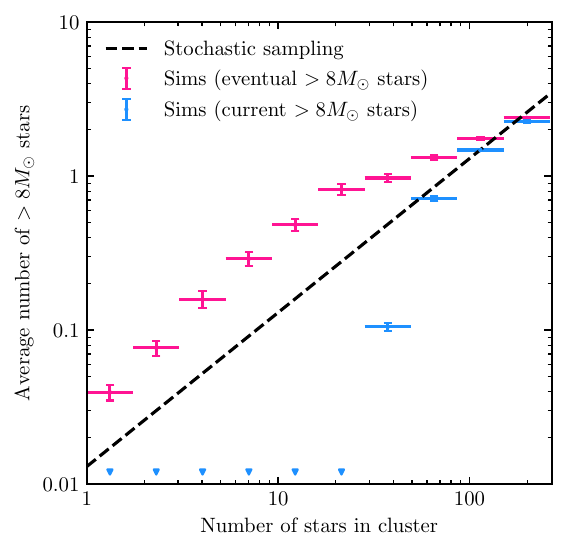}
    \caption{Expected number of massive ($>8M_\odot$) stars in a cluster as a function of the number of stars in a cluster, measured over time in all simulations. We count both the number of stars that will {\it eventually} exceed $8M_\odot$ and the number of stars that {\it currently} exceed $8M_\odot$ at a given number of cluster members. Horizontal bars show bin widths, vertical bars show counting uncertainties, downward arrows indicate bins with no stars. For comparison we plot the expected number of massive stars when drawing from the total sample IMF (Fig. \ref{fig:total_imf}) at random (dashed).}
    \label{fig:Nmassive_expected}
\end{figure}

The results, binned over the full set of simulation snapshots (over all stages of cluster formation), are summarized in Figure \ref{fig:Nmassive_expected}. The average number of eventually-massive stars tends to be $\sim 3\times$ greater than that expected from random sampling -- i.e., massive stars in the simulations tend to {\it start} forming earlier than expected if stars of different masses formed in random order. The opposite is true for the counts of presently-massive stars: there tend to be fewer than expected from random sampling as the cluster forms, while it has $\lesssim  100$ stars. Stars of different masses do not finish forming (i.e., reach their maximum mass) at random times. A massive star is never among the first 20 fully-formed stars in any of these simulations. This is likely due to the finite formation time required for the massive stars to accrete their mass, which can be as long as a few Myr \citep{starforge.fullphysics}, comparable to the the total lifetime of the simulated clouds.



\subsection{Final stellar demographics}

\begin{figure*}
    \centering
    \includegraphics[width=\columnwidth]{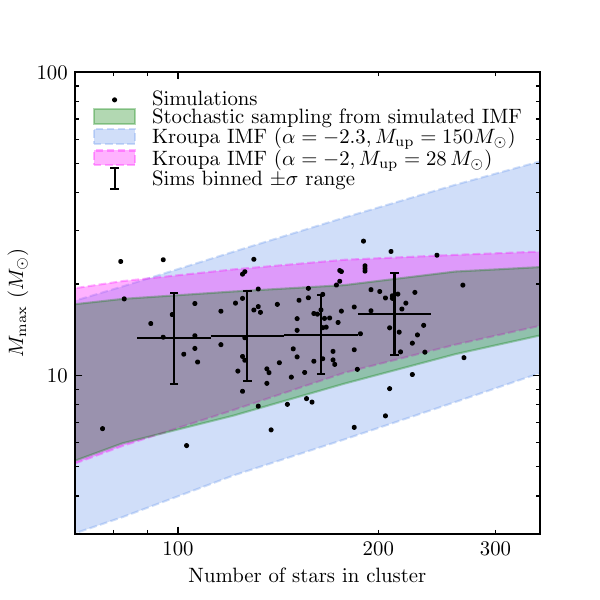}\includegraphics[width=\columnwidth]{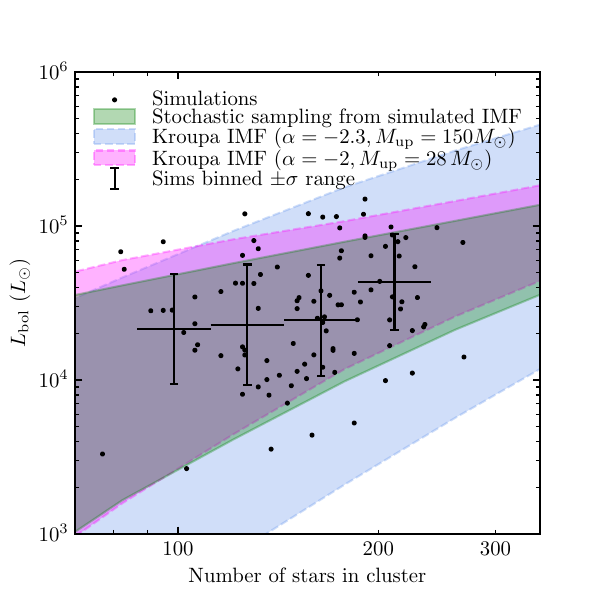} 
    \caption{Relation between number of stars and maximum stellar mass (left) and bolometric luminosity (right) at the end of star formation in the 100 simulations; each datapoint represents one simulated star cluster. Error bars plot the binned median and $\pm \sigma$ range of the simulation results (horizontal bars show bin widths). For comparison we plot the $\pm \sigma$ quantiles of clusters randomly sampled from sample IMF, a \citet{kroupa_imf} with standard slope $\alpha=-2.3$ and cutoff $M_{\rm up}=150M_\odot$, and a modified Kroupa IMF with slope $\alpha=-2$ and hard cutoff $M_{\rm up}=28M_\odot$.}
    \label{fig:mass_to_light}
\end{figure*}

Next we examine the statistics of the most important properties for the fully-formed cluster's protometric and feedback properties: the maximum stellar mass $M_{\rm max}$ and the cluster luminosity $L_{\rm bol}$. In Figure \ref{fig:mass_to_light} we plot the relation between the number of member stars and $L_{\rm bol}$ and $M_{\rm max}$ respectively for the simulated clusters. Note that the two plots are qualitatively identical because $L_{\rm bol}$ is strongly correlated with $M_{\rm max}$ due to the steep scaling of stellar luminosity with stellar mass.

The $L_{\rm bol}$ and $M_{\rm max}$ relations exhibit $\sim 0.4 \rm dex$ and $\sim 0.15 \rm dex$ of scatter, respectively; the number of stars in the cluster does not uniquely predict either quantity. Remarkably, the degree of scatter agrees well with that predicted from stochastic sampling from the sample IMF, or from a modified \citet{kroupa_imf} IMF with slope $\alpha=-2$ and hard cutoff $M_{\rm up}=28 M_\odot$. However, a standard IMF with $\alpha=-2.3$ and $M_{\rm up}=150M_\odot$ exhibits significantly more scatter in both directions due to the different shape and cutoff. 

It is worth re-emphasizing why this result is not necessarily expected: {\it a priori}, the data plotted in Figure \ref{fig:mass_to_light} did not have to exhibit any significant scatter whatsoever, let alone resemble random sampling. In principle star formation could have proceeded as a completely well-determined, organized process in which the macroscopic system properties tightly constrain the outcome \citep[e.g.][]{weidner.kroupa:2006}. The highly-nonlinear, dissipative physics at play might have damped out any particulars of the initial seed and regressed toward an emergent attractor behavior, perhaps a detailed self-regulation by feedback. But this is not the outcome realized by the physics of star formation, as captured by the simulations.

\section{Discussion}

\subsection{The variable high-mass truncation of IMF}
What does it mean that the most-massive known stars in the Local Group tend to be found in the most-massive young star clusters? Is it a size-of-sample effect from sampling a universal power-law \citep{larson_1982}, or do certain clusters form under special conditions that promote the formation of very massive stars? Distinct high-mass truncations are evident in the raw present-day mass functions of 30 Dor and Carina OB2 \citep{2004MNRAS.348..187W,2010ApJ...713..871W}, but with the severe caveat that the present-day mass function of these $>4\, \rm Myr$ old systems will also be truncated from the death of the earliest-forming massive stars. This sensitivity to the star formation history introduces some nontrivial model-dependence to the measurement in evolved clusters and associations. Nevertheless, various observational works restricted to $<4 \rm Myr$ old clusters have found evidence against the pure size-of-sample explanation \citep{2012ApJ...752...59H,2017ApJ...834...94S, 2023A&A...670A.151Y}.



The situation in the simulations is clearer. In \S\ref{sec:imf_truncation} we showed that our 100 low-mass cluster formation simulations, taken together with previous {\small STARFORGE} simulations of other conditions, suggest an environmentally-varying high-mass truncation in the IMF that is generally distinct from the usually-assumed universal $100-150M_\odot$ cutoff. We do not yet have sufficient coverage of parameter space to map out the detailed scaling, but the $10\times$ more-massive cloud IMF shown in  Figure \ref{fig:total_imf} indicates that at least total cloud mass is a significant factor. 



One factor affecting the maximum stellar mass in a given cloud could simply be the time that a star has to draw gas from the cloud. These simulated clouds tend to disperse due to feedback after a time proportional to the free-fall timescale $t_{\rm ff}$, and at fixed mean surface density $t_{\rm ff} \propto M^{1/4}$, so more-massive clouds tend to live longer, and more-extended accretion is possible. At first glance this explanation seems tempting for the cases shown in Fig. \ref{fig:total_imf}, because the $\sim 10 \times$ more-massive cloud produces a star $\sim 10^{1/4}=1.8$ times more massive.

But the cloud lifetime alone does not explain all variation, and predicts the wrong trend in other cases: \citet{guszejnov:2022.starforge.imf} found their densest, shortest-lived cloud produced more-massive stars than the fiducial case. The other facet of the problem is feedback: simulations of these same clouds without feedback were easily able to produce 100+$M_\odot$ stars \citep{guszejnov:2020.isothermal.mhd,starforge_jets_imf}. The relation between the cutoff and extensive cloud properties may therefore be understood in terms of the way massive stars tend to form in these simulations: from accretion of extended gas flows over timescales as long as several Myr \citep{starforge.fullphysics}, which are hardly separated in length- or time-scale from the parent cloud size and crossing time. The accretion of massive stars and the resulting influence of feedback on cloud dynamics are therefore coupled. Denser clouds are generally more resistant to feedback, and may therefore produce more-massive stars if all else is equal.

\subsubsection{Analytic estimates}
To build some analytic intuition for how feedback and cloud properties might influence the upper cutoff of the IMF, let us consider how the stellar mass required to disrupt a cloud might scale under some simplifying assumptions. We estimate an upper bound on $M_{\rm up}$ by determining the mass that a star must have to single-handedly disrupt a cloud within some fraction of a freefall time $N_{\rm ff} \lesssim 1$, i.e. soon enough to prevent significant additional star formation.

If the cloud is disrupted by ionizing radiation, we may adopt the \citet{hosokawa_2006_hii} D-type HII region expansion solution and ask what H ionizing photon production rate $\mathcal{Q}\left(H^{\rm 0}\right)$ is required for the bubble to sweep out the cloud radius $R$ within $N_{\rm ff}$ freefall times, $t_{\rm ff} = \sqrt{3\uppi/\left(32 G\rho\right)}$. Assuming $R$ is much larger than the Str\"{o}mgren radius (and hence the self-similar $\propto t^{4/7}$ expansion law applies), we obtain
\begin{equation}
    \mathcal{Q}\left(H^{\rm 0}\right) \gtrsim \frac{6912 G^2 M_{\rm cl}^{3/2} m_{\rm p}^2 X_{\rm H}^2 \alpha_{\rm B} \left(T_{\rm i}\right)\mu_{\rm i}^2 \Sigma^{5/2}}{2401 \uppi^{5/2} N_{\rm ff}^{4} k_{\rm B}^2 T_{\rm i}^{2}},
\end{equation}
where $M_{\rm cl}$ is the mass of the cloud, $X_{\rm H}$ the mass fraction of H, $\mu_{\rm i} \sim 0.61$ the mean molecular weight of ionized gas, $\Sigma=M_{\rm cl}/\uppi R^2$ the mean cloud surface density, $T_{\rm i} \sim 10^4 \rm K$ the temperature of ionized gas, and $\alpha_{\rm B}\left(T_{\rm i}\right) \approx 2.6 \times 10^{-13} \left(T_{\rm i}/10^4 \rm K\right)^{-0.85} \,\rm cm^3 \,s^{-1}$ is the case B recombination coefficient \citep[][Table 2.1]{osterbrock}. More conveniently,
\begin{equation}
    \mathcal{Q}\left(H^{\rm 0}\right)\gtrsim 10^{47}\,\mathrm{s}^{-1} N_{\rm ff}^{-4} \Sigma_{\rm 2}^{5/2} M_{\rm 4}^{3/2}T_{\rm i,4}^{-2.85},
    \label{eq:QH0}
\end{equation}
where $\Sigma_{\rm 2}=\Sigma/10^2 M_\odot \rm pc^{-2}$, $M_{\rm 4}=M_{\rm cl}/10^4 M_\odot$, and $T_{\rm i, 4} = T/10^4 \rm K$. Approximating the \citet{diazmiller:1998.ionizing.flux} scaling of ionizing flux with stellar mass\footnote{In identifying $\mathcal{Q}\left(H^{\rm 0}\right)$ with a certain stellar mass we have assumed that all of the ionizing flux originates in the most massive star. In practice, the most massive star accounts for $>50\%$ of the total ionizing flux in 95 of our 100 simulations, so this is not an especially bad assumption. In the stellar mass range $\sim 10-30 M_\odot$ $\mathcal{Q}\left(H^{\rm 0}\right)$ scales so steeply with stellar mass that the most massive star tends to dominate.} over the range $10-30 M_\odot$ as a power-law $\log \mathcal{Q}\left(H^{\rm 0}\right) = 44.7 + 7.3 \log\left(M_\star/10M_\odot\right)$, we find
\begin{equation}
    M_{\star} \gtrsim 16M_\odot N_{\rm ff}^{-0.55} \Sigma_{\rm 2}^{0.34} M_{\rm 4}^{0.2} T_{\rm i,4}^{-0.39}
    \label{eq:mstar}
\end{equation}

Hence the maximum stellar mass scales within increasing surface density and cloud mass. For the 100 simulations with $M_{\rm 4}=0.2$ and $\Sigma_{\rm 2}=0.7$, the median maximum stellar mass was $\sim 18 M_\odot$, implying $N_{\rm ff}\sim 0.4$, fairly consistent with the cloud disruption timescales documented for these simulations \citep{starforge.fullphysics,guszejnov:2022.starforge.imf}. Also notable is the dual role of $T_{\rm i}$, setting both the gas pressure driving bubble expansion and the efficiency of recombination. Dramatically lower ionizing fluxes are required to disrupt a cloud if $T_{\rm i}$ is greater, e.g. in very low-metallicity clouds lacking metal line coolants. This agrees qualitatively with the $0.01 Z_\odot$ simulations presented in \citet{guszejnov:2022.starforge.imf}, which did generally have significantly lower maximum stellar masses (and star formation efficiency) than their $Z_\odot$ counterparts.\footnote{This result of lower $M_{\rm max}$ at low metallicity runs counter to the usual intuition that stellar masses should be higher due to suppression of fragmentation -- however, as argued here, this result {\it is} expected if photoionization is the dominant feedback and $M_{\rm max}$ is set by the ability of feedback to regulate accretion, rather than fragmentation physics.}

Photoionization feedback is not universally effective across all of cloud parameter space, so other mechanisms could also affect the maximum stellar mass. The picture may change qualitatively when scaling to sufficiently massive (e.g. $\gtrsim 10^6 M_\odot$) clouds, with freefall times exceeding $10 \rm Myr$, both because the greater escape speed relative to the ionized gas sound speed will make a given amount of feedback less efficient, and because massive stars can die over this longer timescale. 

Consider now the \citet{steigman_1975_momentum_similarity} solution for an expanding bubble driven by a radial momentum source $\dot{P}$: $R=\left(3 \dot{P}/2\uppi \rho\right)^{1/4}t^{1/2}$. This could model direct (single-scattered) radiation pressure on dust grains, or a radiatively-efficient stellar wind bubble. The required $\dot{P}$ to disrupt a cloud or clump is
\begin{equation}
    \dot{P} \geq \frac{4 G M \Sigma}{\uppi N_{\rm ff}^2} = 2.8 \times 10^5  \Sigma_{\rm 2} M_{\rm 4} N_{\rm ff}^{-2} \frac{L_\odot}{c}.
    \label{eq:pdot}
\end{equation}

For photon momentum from a massive star we have 
\begin{equation}    
\begin{split}
    \dot{P} &=\frac{ (1-f_{\rm esc}) L_{\rm bol}}{c}\\ &\approx (1-f_{\rm esc}) \min\left(1.4 \left(M_\star/M_\odot\right)^{7/2}, 3\times 10^4 M_\star/M_\odot \right) \frac{L_\odot}{c},
\end{split}
\end{equation}
assuming a photon escape fraction $f_{\rm esc}$, so Eq. \ref{eq:pdot} gives
\begin{equation}
    \frac{M_\star}{10 M_\odot} \gtrsim \max \left( \frac{ M_{\rm 4}\Sigma_{\rm 2} }{(1-f_{\rm esc})  N_{\rm ff}^{2}},  3\left(\frac{M_{\rm 4} \Sigma_{\rm 2}}{(1-f _{\rm esc})  N_{\rm ff}^{2}}\right)^{2/7}\right),
\end{equation}
so again the maximum stellar mass scales positively with cloud mass and surface density, and the numerical value may plausibly account for the upper IMF cutoff we find for $M_{\rm 4}=0.2$, $\Sigma_{\rm 2}=0.7$ at the order-of-magnitude level, assuming that $(1-f_{\rm esc}) N_{\rm ff}^2 \approx 1$.  

For plausible values of $f_{\rm esc}$ and metallicity, the momentum present in main-sequence OB stellar winds is generally less than that of photons, so the corresponding expression for stellar wind momentum predicts a much higher stellar mass. Stellar winds could be more relevant if they instead maintained a radiatively-inefficient interior \citep{weaver_1977_winds}, but this appears difficult to realize in turbulent GMCs due to venting and strong mixing at the bubble interface \citep{lancaster:2021.wind.sims}. 

 Lastly, as stars approach the Eddington limit ($\gtrsim 30 M_\odot$) multiple-scattered infrared radiation pressure should also become important \citep{larson:1971.masslimit,wolfire:1987.massive.sf,kuiper_2010_massive_sf}, likely tapering off the scaling law in a manner that is sensitive to dust properties. But how the bound on the maximum stellar mass from IR radiation pressure connects to the large-scale cloud properties is less clear, because it operates mainly on much smaller scales: within the dense protostellar envelope, just beyond the dust destruction front at a radius of a few AU.

\subsubsection{Broader implications}
If different cloud properties imprint different upper truncations in the IMF, then there are important implications for the stellar populations of galaxies. GMCs span a broad range of masses and densities within even a single galaxy \citep{heyer:2015.gmcs,freeman:2017.m83.gmcs,sun:2018.gmcs}, so the GMC-level IMFs would vary in their upper truncation and the resulting galaxy-wide IMF would have a different shape. In particular, if $M_{\rm up}$ correlates positively with the host cloud mass then the galactic IMF will generally be steeper than the individual cluster IMFs. This is one way in which our models may be able to reproduce the observed high-mass slope in the field population, which is probably steeper than our measured $\alpha=-2$. This is also a key prediction of the integrated galactic IMF (IGIMF) theory developed by \citet{weidner.kroupa:2006} and many subsequent works (see \citealt{kroupa_2013_imf_review} for review), and it has been argued that it could explain differences between FUV and $\rm H\alpha$ brightness profiles \citep{2008Natur.455..641P}. 

However, our eventual star cluster demographics are more random than e.g. the \citet{kroupa_2013_imf_review} ``optimal sampling" model. Crucially, the simulations have a scatter of $0.13\, \rm dex$ in the relation between cluster mass and the most-massive star, whereas optimal sampling has none. This variation compounds to the $\sim 0.4\, \rm dex$ of scatter in luminosity shown in Figure \ref{fig:mass_to_light}. Nevertheless, it is the systematic variation in $M_{\rm up}$ -- and not so much the specifics of the sampling process -- that is required for the composite IMF to have a different shape from the cluster-level IMF, and our simulations give physical motivation to this hypothesis. Whether the corrections to normal population synthesis are large or small in practice remains to be seen, as it will depend on how specifically $M_{\rm up}$ varies across the vast parameter space of star-forming cloud properties.

Because the fully-formed clusters' photometric properties are similar to stochastic sampling from a truncated IMF in the simulations, our result is similar to the one tested observationally by \citet{andrews:2013.imf.sampling}. They compared photometric models stochastically sampled from a \citet{kroupa:imf} IMF assuming $M_{\rm up}=30M_\odot$ and $120M_\odot$ respectively, finding that $M_{\rm up}=120M_\odot$ is better able to account for the photometry of $\sim 10^3 M_\odot$ young star clusters in NGC 4214\footnote{See however concerns raised by \citet{2014MNRAS.441.3348W}.}. However, we emphasize that 1. our clusters are an order of magnitude less massive, and 2. the simulations suggest that the $M_{\rm up}$ should vary with the properties of the cluster or its progenitor cloud, and $\alpha$ can also vary from the exact \citet{salpeter_slope} value. Thus it is not surprising that a model with a hard cutoff of $30M_\odot$ with $\alpha=-2.35$ exactly across the entire galaxy is in tension with the data. Given the widely-varying properties of giant molecular clouds, we expect there should be conditions in most galaxies capable of producing stars more massive than $30 M_\odot$. 

In principle, star cluster photometry may constrain the variation of $\alpha$ and/or $M_{\rm up}$ in greater detail, e.g. by incorporating a model for their scaling and intrinsic scatter in a full Bayesian forward-model \citep[e.g.][]{krumholz:2019.slug.model}. However in practice $\alpha$ and $M_{\rm up}$ tend to be degenerate because both parameters mainly affect the ratio of ionizing (and hence $\rm H \alpha$) luminosity to bolometric luminosity. Photometric models allowing even $\alpha$ alone to vary freely tend to be degenerate \citep{ashworth:2017.legus.slug}, so simulations may be useful for establishing physically-motivated priors to break degeneracies.
  
\subsection{Implications for simulations and feedback modeling}
Stochastic sampling from a fixed standard IMF has become increasingly popular in numerical simulations of galaxies and star cluster formation, with the aim of modeling feedback, stellar dynamics, and chemical enrichment more accurately on small scales. But in \S\ref{sec:imf_truncation} we found that the conditions we simulated do not allow the formation of $\gtrsim 30 M_\odot$ stars, requiring an environmentally-dependent upper IMF truncation to account for the range of stellar masses observed. Ours is likely an extreme case for very low-mass clusters, but nevertheless the variation of $M_{\rm up}$ could be important for feedback modeling in general \citep{2005ApJ...625..754W,2021A&A...655A..19Y,2023MNRAS.526.1713S}. Fortunately the specific choice of $M_{\rm up}$ will not have a large effect on the rate of core-collapse supernovae from a given stellar population, because the number of progenitors is dominated by$\sim 8M_\odot$ stars, well below the cutoff seen even in these low-mass clusters. Therefore the overall energy budget of supernova feedback will not be highly sensitive if .

But a lower value of $M_{\rm up}$ {\it would} produce significantly less radiative feedback from photoionization and early-time radiation pressure. Compared to a \citet{kroupa:imf} IMF with $\alpha=-2.3$ and $M_{\rm up}=150M_\odot$, a cluster with our sample IMF with $\alpha=-2$ and $M_{\rm up}\sim 28M_\odot$ would have a zero-age bolometric luminosity $5\times$ less, and produce $10\times$ fewer H-ionizing photons. There would also be significantly fewer stars producing long-lived, powerful Wolf-Rayet winds \citep{meynet_2005_wolfrayet}. Since these early feedback mechanisms are key for regulating star formation on cloud and star cluster scales, cloud and star cluster properties and the pre-processing of the ISM for supernova explosions may be affected. Variations in the relative abundances of SN progenitors of different masses would also result in measurably-different chemical abundance patterns from supernova ejecta \citep[e.g.][]{mcwilliam:2013.imf.dwarfs}, and different mass-metallicity relations \citep{2007MNRAS.375..673K,2021A&A...655A..19Y}.

Even if the correct IMF is known, we also found in \S\ref{sec:sfsequence} that the way in which stars of different masses are sampled as the cluster formed is not well-modeled by random sampling. The discrepancy between the number of currently-massive stars and random sampling is partly due to the non-negligible time required for massive stars to form. Hence it is possible that stochastic models could be augmented with a model accounting for the finite time required to assemble a star. Without accounting for the time delay, massive stars would be predicted to form too early in a cluster's lifetime, and their feedback would influence subsequent star formation disproportionately.

\subsection{Caveats}
\label{sec:caveats}
Star formation simulations require many physical approximations and can only treat a segment of the physical dynamic range self-consistently, so their results should be interpreted carefully. A detailed discussion of known and potential caveats and uncertainties of the {\small STARFORGE} numerical model is given in \citet{starforge.methods} \S5.2.

If the scatter in Fig. \ref{fig:mass_to_light} is less than permitted by observations, as argued by \citet{2023A&A...670A.151Y}, then some element is missing from the simulations. Intuitively, if star formation proceeds in roughly a turbulent crossing time \citep[e.g.][]{elmegreen:2000}, as it does in these simulations, then there is insufficient time for different parts of the cloud to communicate, and some of the randomness in the initial conditions will survive. Slowing down the star formation process over multiple crossing times would allow the system to become well-mixed and more self-regulated. In STARFORGE, we see something like this in our most highly-magnetized (mass-to-flux ratio of 0.4) cloud, which has a much smoother, more-quiescent star formation history \citep{guszejnov:2022.starforge.imf}. 

More generally, a potentially-important uncertainty is the initial and boundary conditions. We have simulated an ensemble of isolated, initially-uniform, spherical clouds with Gaussian random initial turbulence, and real GMCs are unlikely to be well-described by any of these properties. Some studies have found that such spherical models may exhibit realistic properties and star formation dynamics once clumpy substructure has had time to form \citep{reyraposo_2015_gmc_zooms,2023MNRAS.519.6392P}. But any dense cloud must inevitably have some continuity with the converging galactic gas flow that formed it, so isolated cloud evolution is unlikely to capture the full dynamics of star formation. The problem of the IMF and its sampling process should be revisited with alternative initial/boundary conditions setups \citep[e.g.,][]{lane:2021.turbsphere}, and ideally a larger-scale setup embedded within a galaxy.

And lastly, we re-emphasize that we have intentionally simulated a very particular case of low-mass star formation to stress-test the usual assumptions of stellar population synthesis, and that this is not representative of the massive complexes that form most stars. We have shown that discrepancies with stochastic sampling exist in detail, and this should not change qualitatively with the scale of the star-forming region. But it remains to be seen whether the differences from stochastic sampling are large in general. For instance, if the sampling always regresses to the mean once $\sim 100$ stars have formed then the correction would be negligible in any massive star-forming complex. But 100 just happens to be the typical cluster size here, so it could also be that clusters regress to the mean on Fig. \ref{fig:Nmassive_expected} only once most stars have formed, in which case the difference from stochastic sampling will be important at all scales. It will be possible to clarify this issue with an upcoming statistics suite of larger clouds, currently in progress.

\section{Conclusion}
In this work we use numerical simulations to test two common assumptions when modeling stellar populations: that stellar masses are randomly, independently sampled, and that the underling IMF is invariant. Within our models, both assumptions break down to some extent:
\begin{itemize}
    \item There is a high-mass truncation in the IMF $M_{\rm up}$ that can generally be significantly lower than largest known stellar masses (Fig. \ref{fig:total_imf}), at least in low-mass clusters. This truncation varies with cloud properties such as mass, so different environmental conditions are required to account for the full observed range of stellar masses, and in general it is not valid to assume that all clouds can form arbitrarily-massive stars.
    \item Stars of different masses do not form in a random order. Massive stars in the simulated clouds begin forming earlier and finish forming later than average, so assuming random sampling during star formation 
    overestimates the number of massive stars that exist at a given time.
\end{itemize}


Although star formation is not random and uncorrelated in our deterministic simulations, stochastic sampling {\it does} appear to provide an adequate description of the {\it end result} of star formation, with the major caveat that the high-mass slope and truncation for the system in question must be known, or generally allowed to vary as a model parameter. Stars do not form through a roll of the dice, but the laws of physics may conspire to make it appear so.

\section*{Acknowledgements}
We thank Todd Thompson, Zhiqiang Yan, Bruce Elmegreen, Martijn Wilhelm, Daniela Calzetti, and Pavel Kroupa for enlightening exchanges that motivated this work. Support for MYG was provided by NASA through the NASA Hubble Fellowship grant \#HST-HF2-51479 awarded  by  the  Space  Telescope  Science  Institute,  which  is  operated  by  the   Association  of  Universities  for  Research  in  Astronomy,  Inc.,  for  NASA,  under  contract NAS5-26555. SSRO was supported by NSF through CAREER award 1748571, AST-2107340 and AST-2107942,  by NASA through grants 80NSSC20K0507 and 80NSSC23K0476 and by the Oden Institute through a Moncrief Grand Challenge award. CAFG was supported by NSF through grants AST-2108230  and CAREER award AST-1652522; by NASA through grants 17-ATP17-0067 and 21-ATP21-0036; by STScI through grant HST-GO-16730.016-A; by CXO through grant TM2-23005X; and by the Research Corporation for Science Advancement through a Cottrell Scholar Award. Support for DG was provided by NASA through the NASA Hubble Fellowship grant \#HST-HF2-51506 awarded  by  the  Space  Telescope  Science  Institute. Support for PFH was provided by NSF Research Grants 1911233, 20009234, 2108318, NSF CAREER grant 1455342, NASA grants 80NSSC18K0562, HST-AR-15800. The suite of 100 new simulations was run on the Anvil supercomputer, through XSEDE award TG-PHY210070. Analysis was performed using Frontera allocation AST21002. This research is part of the Frontera computing project at the Texas Advanced Computing Center. Frontera is made possible by National Science Foundation award OAC-1818253.

\section*{Data Availability}

The complete dataset of simulation snapshots with time-dependendent stellar (sink particle) properties is available at \href{https://data.obs.carnegiescience.edu/starforge/M2e3_R3.tar.gz}{this link}. A public version of the {\small GIZMO} code is available at \url{http://www.tapir.caltech.edu/~phopkins/Site/GIZMO.html}.



\bibliographystyle{mnras}
\bibliography{master}

\end{document}